\documentclass[fleqn,twoside]{article}
\usepackage{espcrc2}
\usepackage{epsfig}
\usepackage{axodraw}
\def\sign(#1){(\!-\!1)^{#1}}
\def\binom(#1,#2){ (\!\!
	 \begin{array}{c} #1 \\ #2 \end{array}\!\! ) }

\def\rb(#1){$#1$}
\def\ob(#1){\framebox[1.1\width]{$#1$}}

\def\bb(#1){$\underline{#1}$}

\newcommand{\AmS}{{\protect\the\textfont2
  A\kern-.1667em\lower.5ex\hbox{M}\kern-.125emS}}

\title{FORM facts}

\author{J.A.M. Vermaseren
        \\ Nikhef, 
        Science Park 105, 1082XG, Amsterdam, The Netherlands}%
\begin{document}

\begin{abstract}
Some of the new features of the symbolic manipulation system FORM are 
discussed. Then some recent results running its multithreaded version TFORM 
are shown. Finally the plans for the future are presented.
\vspace{1pc}
\end{abstract}

\maketitle


\section{New Features}

Over the past few years the symbolic manipulation system FORM\cite{FORM} 
has picked up a number of new features. Some are designed for better speed, 
some for more convenience. The most recent facility is the set of transform 
statements for manipulating functions with a large number of arguments. The 
problem here is that doing the arguments one by one often involves a large 
number of statements, and always a superfluous amount of pattern matching. 
Just imagine we have a function f with 20 arguments. All have a value of 
either zero or one. Now we want to replace the zeroes by ones and the ones 
by zeroes. One way to do this is with
\begin{verbatim}
   Multiply f1;
   repeat id f(x?,?a)*f1(?b) =
                      f(?a)*f1(?b,1-x);
   id f*f1(?a) = f(?a);
\end{verbatim}
Note that the repeat loop has to be done 20 times. With the new transform 
statement we would have

\begin{verbatim}
  Transform,f,replace(1,last)=(1,0,0,1);
\end{verbatim}
On my laptop the first method takes $45.20 \mu$sec. and the second method 
takes $1.38 \mu$sec. In addition the code is easier to understand. This 
gives a smaller chance of errors.

One can have different subkeys and one statement can have a whole chain of 
operations as in
\begin{verbatim}
 CF  H;
 L   F = H(3,4,2,6,1,1,1,2);
 Transform,H,tointegralnotation(1,last),
             replace(1,last)=(0,1,1,0),
             encode(1,last):base=2;
 Print;
 .end

 F =
    H(907202);
\end{verbatim}

One can also split the transform statement in various statements if one 
would like to see what happens:
\begin{verbatim}
 CF  H;
 Off Statistics;
 L   F = H(3,4,2,6,1,1,1,2);
 Print "<1> %t";
 Transform,H,tointegralnotation(1,last);
 Print "<2> %t";
 Transform,H,replace(1,last)=(0,1,1,0);
 Print "<3> %t";
 Transform,H,encode(1,last):base=2;
 Print "<4> %t";
 .end
<1>  + H(3,4,2,6,1,1,1,2)
<2>  + H(0,0,1,0,0,0,1,0,1,0,0,0,0,0,1,1
                               ,1,1,0,1)
<3>  + H(1,1,0,1,1,1,0,1,0,1,1,1,1,1,0,0
                               ,0,0,1,0)
<4>  + H(907202)
\end{verbatim}\vspace{3mm}

The old style program would have been

\begin{verbatim}
   #:WorkSpace 10M
   Symbol x,x1,x2;
   CF  H,H1;
   L   F = H(3,4,2,6,1,1,1,2);
   repeat id H(?a,x?!{0,1},?b) = 
                         H(?a,0,x-1,?b);
   Multiply H1;
   repeat id H(x?,?a)*H1(?b) =
                       H(?a)*H1(?b,1-x);
   id  H*H1(?a) = H(?a);
   repeat id H(x1?,x2?,?a) =
                          H(2*x1+x2,?a);
   Print;
   .end

   F =
      H(907202);
\end{verbatim}
The relevant code here takes $130 \mu$sec. while the composite transform 
statement takes $1.82 \mu$sec. In addition the last code needs commentary 
if one would like to know what it actually does.

The transform statement also allows any permutations of arguments. In the 
example here we have cyclic permutations, the first is one backwards and 
the second is two forwards:
\begin{verbatim}
    CF  f1,f2;
    S   a,b,c,d,e;
    L   F = f1(a,b,c,d,e)*
                    f2(3,2*a,4,c,1,2,3);
    Transform,f1,cycle(1,last)=-1;
    Transform,f2,cycle(1,last)=+2;
    Print;
    .end

   F =
      f1(b,c,d,e,a)*f2(2,3,3,2*a,4,c,1);
\end{verbatim}

There are also facilities for Lyndon\cite{LYND} words of arguments as this is usually 
messy to program externally:
\begin{verbatim}
    Symbol x,x1,x2;
    CF  H,H1,f;
    Off Statistics;
    L   F = H(3,4,2,6,1,1,1,2)
           +H(6,1,1,1,2,3,4,2)
           +H(4,3,2,1,4,3,2,1)
           +H(4,3,2,1,4,2,2,2)
           +H(4,2,2,2,4,3,2,1)
           +H(1,1,1,6,2,4,3,2)
           +H(2,4,3,2,1,1,1,6);
    Transform,H,toLyndon>(1,last)=
                            (f(1),f(0));
    Print +s;
    .end

   F =
       + 2*H(4,3,2,1,4,2,2,2)*f(1)
       + H(4,3,2,1,4,3,2,1)*f(0)
       + 2*H(6,1,1,1,2,3,4,2)*f(1)
       + 2*H(6,2,4,3,2,1,1,1)*f(1)
      ;
\end{verbatim}
The term is multiplied by \verb:f(1): when it is a Lyndon word and by 
\verb:f(0): when it is not. If we would have put just \verb:(1,0): at the 
end of the transform statement the non-Lyndon term would have been absent 
in the output.

The same program, but now with the ordering `smallest first':
\begin{verbatim}
    Symbol x,x1,x2;
    CF  H,H1,f;
    Off Statistics;
    L   F = H(3,4,2,6,1,1,1,2)
           +H(6,1,1,1,2,3,4,2)
           +H(4,3,2,1,4,3,2,1)
           +H(4,3,2,1,4,2,2,2)
           +H(4,2,2,2,4,3,2,1)
           +H(1,1,1,6,2,4,3,2)
           +H(2,4,3,2,1,1,1,6);
    Transform,H,toLyndon<(1,last)=
                            (f(1),f(0));
    Print +s;
    .end

   F =
       + 2*H(1,1,1,2,3,4,2,6)*f(1)
       + 2*H(1,1,1,6,2,4,3,2)*f(1)
       + 2*H(1,4,2,2,2,4,3,2)*f(1)
       + H(1,4,3,2,1,4,3,2)*f(0)
      ;
\end{verbatim}

Large runs suffer from the problem that the computer may not be up that 
long. Jens Vollinga has worked at a checkpoint facility which allows the 
user to make `snapshots' before the start of a module. If FORM crashes, for 
instance due to a power outage, one can restart at the beginning of that 
module.

Currently this is still being debugged and tuned. For some applications it 
is still too slow. There are also still some childhood diseases, but things 
improve.

TFORM\cite{TFORM} has been improved a bit. The master needs far less time 
when there are brackets and the brackets have been indexed. In that case 
the master can tell the workers to deal with complete brackets. From that 
point on each worker is responsible for finding the terms of the brackets 
on its own. In the old setup, the master has to read all terms and put them 
in the `buckets', before giving the buckets to the workers. The speedup is 
noticeable. This still leaves the bottlenecks at the end of the sorting. 
There exist algorithms that might be able to deal with this but they are 
rather complicated. They are planned for a future upgrade.

Last but not least: there have been numerous bug fixes. For this many 
thanks to the people who provide me with concise bug reports that allow me 
to catch these bugs.


\section{Something to boast about}

One of the great testjobs during the development over the past few years 
has been the expression of Multiple Zeta Values\cite{ZAG1,Summer,MZV} in terms of a minimal 
basis. This is mainly a matter of solving a system of linear equations in 
which the coefficients in the homogeneous part are rational numbers and the 
inhomogeneous part contains sums and products of basis elements of a lower 
weight with sometimes rather bad rational coefficients. In the worst 
case the number of equations may run in the millions and the number of 
unknows can be around 1 million or more.

The worst run thus far took 69 days on the 8 cores of one of the nodes of 
the computer in Karlsruhe and verified the conjecture that a new type of 
basis element was going to enter.

One of the statistics in this program:
\begin{verbatim}
Time =   69738.22 sec   Generated terms=
                           6768912520814
              FF        Terms in output=
                              2563910243
substitution(8-sh)-4544 Bytes used     =
                             61564939480
\end{verbatim}
The total number of generated terms in the job was 28,710,904,088,430 which 
is 600,000 terms per second per core.

The number of variables increases with $2^{w-3}$. Before this project was 
started, the mathematicians had gotten to $w=18$ and for $w=19$ and $w=20$ 
they had used matrix techniques to determine only the size of the basis. 
Now we have a full basis up to $w=26$ and $w=28$. For $w=27$ we still miss 
two basis elements but we can guess them.

At the same time we studied something discovered earlier by 
Broadhurst\cite{Broadhurst:1}, 
called pushdowns in which basis elements of the MZV's could be expressed in 
terms of alternating sums with fewer indices as in:
\begin{eqnarray}
    Z_{6,4,1,1} & = &
       - \frac{64}{27} A_{7,5}
       - \frac{7967}{1944} Z_{9,3}
       + \frac{1}{12} \zeta_3^4
        \nonumber \\ &&
       + \frac{11431}{1296} \zeta_7 \zeta_5
       - \frac{799}{72} \zeta_9 \zeta_3
       + 3 \zeta_2 Z_{7,3}
        \nonumber \\ &&
       + \frac{7}{2} \zeta_2 \zeta_5^2
       + 10 \zeta_2 \zeta_7 \zeta_3
       + \frac{3}{5} \zeta_2^2 Z_{5,3}
        \nonumber \\ &&
       - \frac{1}{5} \zeta_2^2 \zeta_5 \zeta_3
       - \frac{18}{35} \zeta_2^3 \zeta_3^2
       - \frac{5607853}{6081075} \zeta_2^6~ \nonumber
\end{eqnarray}
with $A_{7,5} = H_{7,5}-H_{-7,5}$.

In ref~\cite{MZV} we managed to locate 16 of such relations in a combined 
symbolic (FORM) and numerical (PSLQ) effort.

\begin{table*}[h!tb]
\begin{center}
\begin{tabular}{|c|c|c|c|c|c|c|c|c|c|c|}
\hline
  {\sf w/d}& 1& 2 &  3    &  4     &  5     &  6       &  7       &  8        &  9    & 10 \\ 
\hline
  1 &      &      &       &        &        &          &          &           &       &  \\
  2 &\rb(1)&      &       &        &        &          &          &           &       &  \\
  3 &\rb(1)&      &       &        &        &          &          &           &       &  \\
  4 &      &      &       &        &        &          &          &           &       &  \\
  5 &\rb(1)&      &       &        &        &          &          &           &       &  \\
  6 &      &\rb(0)&       &        &        &          &          &           &       &  \\
  7 &\rb(1)&      &       &        &        &          &          &           &       &  \\
  8 &      &\rb(1)&       &        &        &          &          &           &       &  \\
  9 &\rb(1)&      &\rb(0) &        &        &          &          &           &       &  \\
 10 &      &\rb(1)&       &        &        &          &          &           &       &  \\
 11 &\rb(1)&      &\rb(1) &        &        &          &          &           &       &  \\
 12 &      &\rb(1)&       &\rb(0,1)&        &          &          &           &       &  \\
 13 &\rb(1)&      &\rb(2) &        &        &          &          &           &       &  \\
 14 &      &\rb(2)&       &\rb(1)  &        &          &          &           &       &  \\
 15 &\rb(1)&      &\rb(2) &        &\rb(0,1)&          &          &           &       &  \\
 16 &      &\rb(2)&       &\rb(2,1)&        &          &          &           &       &  \\
 17 &\rb(1)&      &\rb(4) &        &\rb(1,1)&          &          &           &       &  \\
 18 &      &\rb(2)&       &\rb(4,1)&        &\rb(0,1)  &          &           &       &  \\
 19 &\rb(1)&      &\rb(5) &        &\rb(3,2)&          &          &           &       &  \\
 20 &      &\rb(3)&       &\rb(6,1)&        &\rb(1,2)  &          &           &       &  \\
 21 &\rb(1)&      &\rb(6) &        &\rb(6,3)&          &\rb(0,1)  &           &       &  \\
 22 &      &\rb(3)&       &\rb(10,1)&       &\bb(3,4)  &          &           &       &  \\
 23 &\rb(1)&      &\rb(8) &        &\bb(11,4)&         &\bb(1,3)  &           &       &  \\
 24 &      &\rb(3)&       &\bb(14,2)&       &\bb(8,6)  &          & \bb(0,1)  &       &  \\
 25 &\rb(1)&      &\rb(10)&        &\bb(18,5)&         &\bb(4,7)  &           &       &  \\
 26 &      &\rb(4)&       &\bb(19,1)&       &\bb(16,11)&          & \bb(1,4)  &       &  \\
 27 &\rb(1)&      &\rb(11)&        &\bb(29,7)&         &\bb(11,12)&           & 0,1,1 &  \\
 28 &      &\rb(4)&       &\bb(25,2)&       &\bb(31,14)&          &\ob(4,11,1)&       &  \\
 29 &\rb(1)&      &\rb(14)&        &\bb(42,8)&         &\bb(25,23)&           & 1,5,1 &  \\
 30 &      &\rb(4)&       &\bb(33,2)&       &\bb(52,21)&          & 14,22,1   &       & 0,1,1 \\
\hline
\end{tabular} \vspace{3mm} \\
{Table 1: Number of MZV basis elements as a function of weight and depth}
\end{center}
\end{table*}

In table 1 we give the numbers of basis elements of a certain type: the 
number of elements belonging to the set of Lyndon words with odd integers 
greater than 1 (and adding up to the weight), the number of such elements 
in which the first two indices have been lowered by one and two ones have 
been added, and finally the number of such elements in which the first four 
elements have been lowered by one and 4 ones have been added. As in
\begin{eqnarray}
                        &  & H_{5,3,5,3,5,3,3} \nonumber \\
  H_{7,5,3,3,3,3,3} & \rightarrow & H_{6,4,3,3,3,3,3,1,1} \nonumber \\
  H_{7,5,7,5,3} & \rightarrow & H_{6,4,6,4,3,1,1,1,1} \nonumber
\end{eqnarray}
In the left and top of table 1 we have verified that the number 
of elements in which the first two indices have been lowered by one and two 
ones have been added corresponds exactly to the number of pushdowns. For 
the underlined numbers we can determine a basis but testing explicit 
pushdowns is beyond our reach. The number in the box corresponds to the new 
run in which we found the new basis element for which the first four 
indices are lowered by one and four indices one have been added. The case 
with $W=27, D=9$ is currently running.

The fact that this system of basis construction and the pushdowns give the 
same numbers is very suggestive.


\section{Future Features}

When we are looking towards the future we should first consider who are 
doing the work. I have compiled a list of people who have and are working 
at FORM during various stages. It is shown in table 2.


\begin{table}[h!tb]
\begin{center}
\begin{tabular}{ll}
JV              		 &   1984-now   \\
Geert Jan van Oldenborgh &   manual(90's) \\
Andre Heck     			 &   manual(90's) \\
Albert Retey   			 &   1997-2000 \\
Denny Fliegner 			 &   1998-2000 \\
Markus Frank   			 &   2000      \\
Andrei Onishchenko       &   2000-2002 \\
Misha Tentyukov			 &   2002-now  \\
Jens Vollinga  			 &   2007-now  \\
Thomas Reiter  			 &   2008-now  \\
Irina Pushkina 			 &   2009-now  \\
Jan Kuipers    			 &   2009-now
\end{tabular} \vspace{3mm} \\
{Table 2: People who have worked at FORM}
\end{center}
\end{table}

Then there are of course the beta testers who sometimes put in much work to 
produce a concise bug report or who come up with useful suggestions. A 
number of the most important ones are Ettore Remiddi, Kostia Chetyrkin, 
York Schr\"oder, Thomas Hahn, Takahiro Ueda and Peter Uwer. My apologies if 
I forget people here.

 
\subsection{Open Source}

Sometimes one would like to have quick private additions for things that 
are extremely hard to program at the FORM level. Such things are often 
either of combinatoric nature or special patterns. It is of course 
impossible to forsee what some people will need. Hence FORM should be 
structured in such a way that it is possible to make such additions 
oneself, even though this won't be for beginners. The first requirement for 
this is a good documentation of the inner workings, including a number of 
examples. The second requirement is code that can be understood and is 
structured properly. Due to these two requirements FORM hasn't been 
released yet as open source. We hope to be rectify this by this summer. 
Jens Vollinga is working hard at it.


\subsection{Rational Polynomials}

Systems of equations that need to be solved are asking often for 
capabilities with rational polynomials. Most notoriously are the 
Laporta\cite{Laporta:1,Laporta:2} algorithms. This is something that FORM 
doesn't have currently. Hence it has rather high priority to build this in. 
And to build this in in a rather efficient way as belonging to FORM. There 
exist libraries for the manipulation of polynomials in a single variable, 
some of them claiming great efficiency, but there are no equivalent 
libraries for polynomials in many variables. In addition there is the 
problem of notation. Too much time spent on conversion will not be 
beneficial. Some partial code exists. Most univariate algorithms (in 
particular the GCD) have been implemented in various methods. This is by 
now reasonably fast. Factorization is completely missing. Jan Kuipers is 
working on this and also the multivariate cases.

It is important to deal with multivariate rational polynomials efficiently 
when one likes to create a system for computing Gr\"obner bases. There are 
however several ways to deal with polynomials and each way needs its own 
solution:
\begin{itemize}
\item Small polynomials: when they take a small amount 
of space they can be kept inside the argument of a function. There may be 
billions of such polynomials. They should be treated inside the regular 
workspace. Univariate polynomials will usually be in this category.
\item Intermediate polynomials: these could be 
handled by means of memory allocations as is done with the dollar 
variables. One could have hundreds or even thousands of them. Typically not 
billions.
\item Large polynomials: These are complete 
expressions that could have billions of terms. Calculating their GCD would 
have to use the same mechanisms by which expressions are treated. There 
should be only very few of these.
\end{itemize}

\begin{verbatim}
  Symbols x,y;
  CFunction pacc;
  PolyRatFun pacc;
  L  F=pacc(x^2+x-3,(x+1)*(x+2))*y
   pacc(x^2+3*x+1,(x+3)*(x+2))*y^2
       ;
  Print +s;
  .sort

 F =
    +y*pacc(x^2+x-3,x^2+3*x+2)
    +y^2*pacc(x^2+3*x+1,x^2+5*x+6)
    ;

  id  y = 1;
  Print;
  .end

 F =
    pacc(2*x^2+4*x-4,x^2+4*x+3);
\end{verbatim}

 
\subsection{Code Simplification}

We like to have a way to introduce code 
simplification. This would be relevant for all outputs that would need 
further numerical evaluation in the languages Fortran and C. If it is 
possible we would like to extend this to the regular output for as far as 
factorization is concerned. Already some things can be done at the FORM 
level, but this is usually rather slow. One can for instance make a 
procedure `tryfactor' which would work like

\begin{verbatim}
    #do i = -100,100
    #call tryfactor(acc,x+`i')
    #enddo
    B acc;
    Print;
\end{verbatim}

and the answer might be like

\begin{verbatim}
    +acc(x-27)*acc(x+6)*acc(x+67)*
                         (.......)
\end{verbatim}

This is however far from ideal. Irina Pushkina is working on improving 
things here and providing internal code for such simplification.

 
\subsection{ParFORM x TFORM}

The ParFORM\cite{ParForm} subproject of the Sonderforschungsbereich project 
in Karlsruhe is coming to a close. This was lately worked at by Misha 
Tentyukov. We are considering asking for new funds to combine the 
techniques of ParFORM and TFORM, so that we can obtain efficient running on 
clusters of multicore machines. One example of such a computer is the 
Silicon Graphics computer at Karlsruhe which has 24 nodes, each with 8 
cores and its own hard disk of 4 Tbytes.

This is still in the planning stage.

\vspace*{5mm}
\noindent
{\bf Acknowledgments}

The work has been supported by the research program of the Dutch
Foundation for Fundamental Research of Matter (FOM). Many thanks go 
to the university of Karlsruhe and the DFG for the use of their computer.


\end{document}